\documentclass[sigconf, screen, nonacm]{acmart}

\AtBeginDocument{%
  }

\usepackage{booktabs}
\usepackage{graphicx}
\usepackage{natbib}
\usepackage{url}
\acmConference[WebSci '25]{ACM Web Science Conference}{May 20--24,
  2025}{New Brunswick, New Jersey, USA}
\begin{document}

\title[A Little Bubble of Friends]{‘A Little Bubble of Friends’: An Analysis of LGBTQ+ Pandemic Experiences Using Reddit Data}


\author{Dhruvee Birla}
\affiliation{%
 \institution{International Institute of Information Technology}
 \city{Hyderabad}
 \state{Telangana}
 \country{India}}

\author{Nazia Akhtar}
\affiliation{%
  \institution{International Institute of Information Technology}
 \city{Hyderabad}
 \state{Telangana}
 \country{India}}



\begin{abstract}
  Social media was one of the most popular forms of communication among young people with digital access during the pandemic. Consequently, crucial debates and discussions about the pandemic crisis have also developed on social media platforms, making them a great primary source to study the experiences of specific groups and communities during the pandemic. This study involved research using LDA topic modeling and sentiment analysis on data obtained from the social media platform Reddit to understand the themes and attitudes in circulation within five subreddits devoted to LGBTQ+ experiences and issues. In the process, we attempt to make sense of the role that Reddit may have played in the lives of LGBTQ+ people who were online during the pandemic, and whether this was marked by any continuities or discontinuities from before the pandemic period.
\end{abstract}




\maketitle


\section{Introduction}
Coronavirus (COVID-19) is a highly infectious disease caused by the severe acute respiratory syndrome coronavirus 2 (SARS-CoV-2). In December 2019, initial cases of coronavirus were detected in China. With its rapid spread to other countries, it became a global outbreak and was declared a pandemic on 11 March 2020 by the World Health Organisation (WHO)\cite{who23}.
 
With the onset of the pandemic in 2020, regulations to check the spread of the disease were put in place in different countries. The main aim of these regulations was to contain the spread of the virus through physical isolation. These measures were implemented differently and to different degrees in different countries and regions, with some choosing total isolation to others who decided not to make significant changes\cite{hiscott2020global}.
 
Previous pandemics and epidemics of infectious diseases have brought in their wake an intensification of gender inequalities in access to healthcare, social support, education, and employment at a global level\cite{wenham2020women,stemple2016human}. The COVID–19 pandemic was no exception to this norm\cite{al2021investigating,gausman2020sex,phillips2020addressing,yuan2023minority,carli2020women,alon2020impact,flor2022quantifying,fish2021sexual}. LGBTQ+ individuals and groups living in different countries also experienced an intensification of discrimination, prejudice, and violence during the COVID–19 pandemic. LGBTQ+ individuals who were in forced proximity to abusive family members had to either conceal their sexuality or, if found, endure physical and emotional harassment\cite{fish2021sexual,gato2020home}. Prior research found violence from intimate partners among LGBTQ+ communities had increased during the pandemic in different locations, leading to added stress and anxiety among such individuals\cite{stephenson2022covid,kumar2021sexual,juwono2024prevalence}. Additionally, the shift to virtual events and gatherings reduced the chance to connect with one's existing support network through regular, in-person activities, such as attending schools and offices, and Pride Celebrations, that provided respite from abusive interactions. Finally, while hospitals worldwide were generally beleaguered by the sheer number of COVID–19 cases and were forced to take resources away from other kinds of healthcare sectors, a specifically gendered problem that intensified the struggles of transgender people were the lack of access to antiretroviral therapy (ART) drugs and hormone supplements for Gender Affirmation surgeries. To make things worse, many countries, such as the USA, enacted laws that added to the stigma against LGBTQ+ people.
 
While there are many other ways to source data to study such experiences and impact  associated with the COVID–19 pandemic among different LGBTQ+ groups, social media data has recently emerged as an important and viable source for computational work in this area. The increase in the use of social media has prompted researchers to study the positive and negative effects of social media on vulnerable communities with digital access, including but not limited to women, adolescents, and LGBTQ+ communities\cite{dhiman2023impact,rosen2022social}. In the case of LGBTQ+ communities, research suggests that social media platforms have provided a medium for LGBTQ+ people to form supportive networks and freely express their identities among like–minded individuals with similar experiences\cite{dhiman2023impact,craig2021can,hiebert2021finding,lucas2022lgbtq+}. However, social media has also exposed LGBTQ+ people to cyberbullying. Online harassment, hate speech, and discrimination significantly impact their mental health and well–being\cite{dhiman2023impact,glaad2024smsi,yurcaba2022socialmedia,intelligent2022lgbtqreport,fitri2019sentiment,hoiriyah2023sentiment}.
 
With the rapid spread of COVID–19, more and more people with digital access turned to social media to gather and share information\cite{goel2020social,bao20202019,mason2021social}. Analyzing the content available on social media platforms can help understand the prevalent public behaviors and attitudes during an outbreak\cite{li2020retrospective}. This article focuses on a study of data from the social media platform Reddit to understand the experiences of LGBTQ+ people during the COVID–19 pandemic. Founded in 2005, Reddit is a semi–anonymous platform allowing users to post submissions and write comments semi-anonymously by creating handles, allowing people to use alternate identities. It is an American social news aggregation, web content rating, and discussion website self-proclaimed as the ``front page of the Internet"\cite{singer2014evolution}. With 430 million monthly active users, it is ranked as the 7th most visited website as of November 2023\cite{semrush2024}\footnote{Semrush is a SaaS platform founded in 2008 and offered by the American company Semrush Holdings, Inc.}. As of January 2024, approximately 40\% of Reddit’s demographic consists of users from the United States, followed by the United Kingdom, India, and Canada\cite{worldpopulationreview2023reddit}\footnote{Our research suggests that this is a fairly stable demographic profile for Reddit over the years. Furthermore, beyond this basic profile, it is difficult to glean from the Subreddits themselves a more granular demographic profile of users, since the Reddit USP and operational mode is essentially “semi-anonymous.”}.

\section{Related Work and Gaps}

As mentioned above, studies of social media using computational methods abound. Data available on social media platforms can be analyzed for several tasks, including but not limited to fake news detection, user behavior analysis, and brand monitoring. To obtain valuable insights from large amounts of data available on social media platforms, it is necessary to meticulously process the data and choose the appropriate tool to perform a comprehensive analysis\cite{gole2015survey}. Approaches include conducting statistical analysis, text mining, opinion mining, and content analysis\cite{tiwari2020social}. 
 
Research in this area has also specifically focused on gendered concerns. \citet{craig2021can} conducted a mixed–methods online survey and explored the benefits of social media for LGBTQ+ youth by developing a Social Media Benefits Scale (SMBS). Adopting a combined qualitative and computational approach, research by \citet{saha2019language} characterized minority stressors in the discussions shared on the r/lgbt Reddit community. There has also been a growing interest in understanding the role of social media in identity exploration and identity formation\cite{lucero2017safe,hatchel2017digital}. Despite the positive impacts of social media, hate speech and online offensive language during the last decade have been identified as a global epidemic, with the consequences being much worse for LGBTQ+ individuals\cite{kumaresan2023homophobia,gao2020offensive}. \citet{kumaresan2023homophobia} created a dataset for detecting homophobia and transphobia in Malayalam and Hindi languages using comments under the videos of famous YouTubers who make videos talking about LGBTQ+ communities in the hopes that their subscribers will be more welcoming towards LGBTQ+ people. Additionally, \citet{khan2022preliminary} performed a macro–level quantitative analysis of LGBTQ+ tweets since 2006 to gather insights into the emotions and toxicity levels prevalent in LGBTQ+ discussions and observed changes in emotions and toxicity levels over time.
 
There is a growing interest in understanding the underlying issues present in digital media and the growing popularity of big data analytics, leading to questions about whether such data ethically represent marginalized communities, such as LGBTQ+ groups\cite{gieseking2018size,ruberg2020data}. Big data is a major emerging concern in queer digital studies\cite{ruberg2020data,zeffiro2019towards,guyan2022fixing,soesanto2023sentiments}. It can be empowering and excluding at the same time, for it opens up new possibilities but also amplifies the already existing inherent biases and harmful features on social media platforms, which result in biased knowledge production, routine exploitation, and erasure of vulnerable communities, including but not limited to people of color, queer, disabled, and non–Western people\cite{tacheva2024challenging,leurs2017feminist}. 

LGBTQ+ individuals and communities have also tried to assert and implement intersectional frames of reference in these contexts, stressing differential experiences within the broad umbrella of “LGBTQ+” as well as the existence of other kinds of oppression—such as race, class, caste, and religion—which can compound oppressions of gender and sexuality. Therefore, to produce meaningful stories and robust analysis that is sensitive to these issues, researchers are shifting towards performing situated, reflexive, and ethically–sound data–driven research, thereby addressing intersectional power relations such as gender, age, location, and class. An important consequence of this shift is the emphasis that data–driven knowledge production is never neutral; it is context–specific, subjective, and power–ridden\cite{leurs2017feminist,bivens2015under}.
  
However, while there has been significant research on these themes using social media data and computational methods, some gaps remain. Firstly, prior research has very rarely performed qualitative analysis of the results obtained from their computational tools, which is essential to obtain meaningful insights. In this study, we manually analyzed a part of our dataset in order to arrive at a nuanced, granular understanding of our quantitative results. Secondly, to understand the intensified difficulties of vulnerable communities during the pandemic, a similar analysis needs to be conducted for the pre–pandemic period in a similar setting without relying on secondary sources to provide this context. This study includes the creation of such a dataset, which allows us to compare experiences and articulations from both before and during the pandemic. Lastly, even though research has been conducted on various social media platforms, we have gathered data from more, i.e. five, different subreddits to ensure the patterns that emerge in results are not outliers but rather a uniform observation. Finally, there are gaps in the interpretation of sentiment analysis results in previous studies, which we have attempted to address and explain here through our quantitative and qualitative analysis. This study, thus, aims to offer a more rigorous engagement with the data and context of a social media platform, in order to make sense of current debates in queer and social media studies about the empowering or disempowering nature of social media platforms for LGBTQ+ voices online.

\section{Research Questions, Objectives, and Argument}

The present study poses and attempts to answer the following questions using computational methods on Reddit data associated with the pandemic:
\begin{itemize}
    \item What discussions emerge from the analysis of LGBTQ+-centric subreddits during the pandemic? More specifically, what insights can these discussions within online public spaces contribute to our understanding of the lived experiences of such users?
    \item What kinds of affect and attitudes emerge from LGBTQ+-centric subreddits during the pandemic? How and why are these similar or different from expressed sentiments, views, and opinions before the pandemic?
    \item What role did social media platforms, such as Reddit, play in the lives of LGBTQ+ people during the pandemic?
\end{itemize}
 
Our objectives, therefore, are as follows:
\begin{enumerate}
    \item To document, collect, and analyze topics from the pandemic period from LGBTQ+-centric subreddits;
    \item To detect, classify, and interpret sentiment from the pandemic period from LGBTQ+-centric subreddits, and compare them with the same from a specified time period before the pandemic; and
    \item By synthesizing insights from topics and sentiments, to arrive at insights about the space offered by these subreddits for LGBTQ+ people during the pandemic and, therefore, the real and potential role of social media in creating safe spaces for LGBTQ+ people.
\end{enumerate}
 
As discussed above, previous research has also shown that already existing biases associated with gender and sexuality worsened during the pandemic and were justified by patriarchal norms and conventions. This paper argues that in this climate of aggravated discrimination and prejudice, Reddit offered a “little bubble of friends” for vulnerable LGBTQ+ people of different backgrounds and in different locations to turn to and communicate their thoughts, views, feelings, and fears. We do not claim that Reddit offered a non–cis–heteronormative utopia, but that its features and the way that users deployed the platform itself ensured that it offered an alternative (virtual) safe haven to exist, articulate, and even thrive in during a particularly hostile time for non–cis–heteronormative expressions and worldviews. In this sense, with all their flaws, these subreddits offer a way of imagining what non–cis–heteronormative spaces—virtual and real—can look and feel like.

\section{Data}

Reddit has over 2.2 million subreddits, which are self-created, independent, theme-based forums created and maintained by their moderators. Registered Reddit users follow these subreddits instead of following other users, therefore being united by a specific topic\cite{medvedev2019anatomy}. Registered users can vote and comment on the posts. The number of votes (including upvotes and downvotes), comments, and the rate of receiving votes are some of the deciding factors in a post's popularity.
 
We chose Reddit because, first, users can express themselves in much more detail owing to the larger character limit of the posts and comments; and second, Reddit is significantly different from other social media platforms in terms of its architecture. It is a semi-anonymous platform allowing for community-based, rather than follower-based, information flow. Arguably, this contributes to a quality discussion around a specific topic without requiring users to reveal their identity or other personal details, which can be a concern among vulnerable people, such as LGBTQ+ groups\cite{treen2022discussion}.
 
To interpret the pandemic experiences of LGBTQ+ people on Reddit, we recovered data from several subreddits. Since we were interested in the COVID–19 pandemic experiences, our chosen time period was 10 March 2020 to 31 August 2021. The reason for choosing these dates was that it was in March 2020 that WHO declared COVID–19 a pandemic, and by the 3rd quarter of 2021, the second wave of COVID–19 was either declining or had ended, and many high-income countries had started offering booster vaccine shots to their residents\cite{world2022director}. We decided to compare our analysis from the pandemic period to the period before the pandemic (January 2019–December 2019) to see if there is a difference or similarity in expression and experiences\footnote{We gathered the ‘before’ pandemic data through subreddit dumps made available by a Reddit user, u/Watchful1. In contrast, data from the pandemic was gathered using a third-party service called Pushshift.}.
 
We obtained data from five LGBTQ+-centric subreddits with English as their primary language and a minimum of 20,000 subscribers. These include r/lgbt, r/LGBTnews, r/LGBTeens, r/ainbow, and r/LGBTQ (see Table \ref{tab:table1})\footnote{While collecting data from before the pandemic period, the dumps did not have data from the r/LGBTQ subreddit. Therefore, our analysis for before the pandemic period is based on the other four subreddits mentioned.}. In Reddit, the posts are of three types: videos, images, and text. Each of these posts has a title. We collected the titles of the posts and the comments made under them. All comments, including one made directly to the post and one that has a reply–to relationship with a comment, are viewed as individual comments and are included in the dataset.
 
Each subreddit offers a description of its purpose, and the descriptions for all five chosen by us for the purpose of this study stress that they were created as safe spaces for LGBTQ+ folks and allies to discuss their lives, share queer issues from varying perspectives, and support each other.

\begin{table}[h!]
\renewcommand{\arraystretch}{1.25} 
\centering
\small
\caption{Subscriber Size of Studied Subreddits (Source: The corresponding subreddit platform)}
\label{tab:table1}
\resizebox{\columnwidth}{!}{%
\begin{tabular}{|c|c|c|}
\toprule
    \textbf{Subreddit} & \textbf{Number of Subscribers (as of January 2024)} & \textbf{Creation Date}\\
    \midrule
    r/lgbt & 1.1M & 14 March 2008\\
    \hline
    r/LGBTnews & 82.1K & 09 October 2009\\
    \hline
    r/LGBTeens & 170K & 17 December 2011\\
    \hline
    r/ainbow & 176K & 12 January 2012\\
    \hline
    r/LGBTQ & 30.1K & 02 September 2009\\
    \bottomrule
\end{tabular}}
\end{table}

Over 2 million comments and approximately 300,000 post titles were obtained from data gathered ‘during’ the pandemic period. In contrast, we gathered approximately 800,000 comments and 100,000 post titles from ‘before’ the pandemic. These numbers also include submissions and comments that were deleted by bots in line with the rules set by the moderators of the respective subreddits.

\section{Methods}

\subsection{Topic Modeling}

LDA topic modeling is a generative probabilistic model of a corpus where the documents are represented as random mixtures over latent topics, and each topic is characterized by a distribution over words. It automatically clusters similar word patterns into different sets, each describing a topic. It efficiently processes large amounts of data into meaningful topics\cite{blei2003latent}. Using this method is effective in understanding the primary areas of concern of LGBTQ+ individuals active in our chosen subreddits.
 
We performed topic modeling for two datasets. The first dataset included all post titles gathered from during the pandemic, and the second included all post titles gathered from before the pandemic. We chose to capture the post's title instead of the content for two reasons. First, the title of the post is tailored to attract attention to the post's content. Therefore, users make it as “attention-grabbing” as possible, which means that the title is a crucial element in the post\cite{weissburg2022judging,medvedev2019anatomy}. Second, titles are mandatory for posting on subreddits, thus making it computationally convenient to gather data.
 
To quantitatively evaluate the performance of our model, we use two metrics, namely, the coherence score using the C\_V coherence model and the perplexity score. To find the best possible coherence score, we had to finetune our hyperparameters, which included alpha, beta, number of iterations, and the number of passes. For instance, creating too few topics results in the overgeneralization of text, whereas generating too many topics results in trivial and overlapping topics.

\subsection{Sentiment Analysis}

For sentiment analysis, we have used the RoBERTa-base model, which was developed by Cardiff University in 2022\cite{loureiro2022timelms}. This method was only performed on the data obtained from the comments of each subreddit. We ignored the post titles because they provided us with no particular verbal opinion on the discussion in question. An example of a post title is: “Canada presents bill banning conversion therapy.” Comments, on the other hand, are often candid perspectives of the users, thus giving us an insight into the opinions of the diverse user base. For example, here is an example of a comment under a post: “I am so so sad. And I am absolutely terrified”.


Unlike our approach to topic modeling, for sentiment analysis, we deemed it best to analyze the sentiments of each subreddit separately since every subreddit sees itself as a specific entity with a different set of rules. This would help us understand the difference in the prevailing sentiment in different subreddits.
 
To finetune our model, annotators labeled 1000 comments as either positive, negative, or neutral. Two hundred comments were randomly generated from each subreddit. Previous research has stressed the importance of high–quality datasets for research on sensitive topics concerning vulnerable groups, which requires annotators and other actors working on the research to be aware of the lived experiences of the communities being studied\cite{kumaresan2023homophobia}. Therefore, three LGBTQ+ annotators labeled the randomly generated comments for our research. The final inter-rater score was 0.59 using Krippendorff’s Alpha. We finetuned the model on the 800 comments and performed a five-fold cross-validation. The test dataset consisted of the remaining 200 comments. This dataset consisted of the majority label for each comment as assigned by the annotators.

Sentiment analysis aids in understanding the attitudes and the overall emotional well–being of individuals. Through our temporal analysis using topic modeling and sentiment analysis, we get insights into the evolving nature of LGBTQ+–related concerns and the variations in their emotional expressions. The results from our chosen methods would provide insights into the influence of the pandemic on the lives of a subpopulation of LGBTQ+ communities. The influence of the pandemic is highlighted through our analysis of data obtained from before the pandemic.

\section{Results}

\subsection{Topic Modeling}

\subsubsection{Time Period I (10 March 2020 – 31 August 2021)}

We found that the most optimal alpha and beta values for this time period were 0.5 and 0.01, respectively. The best number of passes was 50, and the most optimal number of topics was 20. This model's coherence and perplexity scores were 0.42 and -12.66, respectively. From the topic models trained on the optimal hyperparameters, we interpreted the topics based on a combination of 10 keywords and the weight assigned to each (see Table \ref{tab:table2} and Appendix \ref{during}). 

\begin{table}[ht]
  \caption{Major topics during the pandemic period}
  \label{tab:table2}
  \resizebox{.95\columnwidth}{!}{
  \begin{tabular}{|p{2cm}|p{5cm}|p{3cm}|}
    \toprule
    \textbf{Topic Number} & \textbf{Frequent Terms} & \textbf{Inferred Topics} \\
    \midrule
    {1} & {0.123*"pride" + 0.031*"rainbow" + 0.029*"students" +
    0.029*"month" + 0.026*"flag" + 0.019*"protest" + 0.016*"march" +
    0.014*"flags" + 0.013*"lgbtq" + 0.012*"university"} & {Pride
    Celebrations} \\
    \hline
    {2} & {0.064*"biden" + 0.053*"amp" + 0.039*"law" + 0.033*"lgbtq" +
    0.024*"two" + 0.024*"hungary" + 0.023*"joe" + 0.022*"calls" +
    0.021*"day" + 0.018*"president"} & {LGBTQ+--related political
    discussions worldwide} \\
    \hline
    {3} & {0.048*"black" + 0.031*"coronavirus" + 0.026*"say" +
    0.025*"virginia" + 0.024*"love" + 0.023*"lives" + 0.018*"lgbtq" +
    0.018*"new" + 0.017*"show" + 0.017*"years"} & {Stigma against the
    community during the pandemic} \\
  \bottomrule
\end{tabular}
}
\end{table}


\subsubsection{Time Period II (January 2019 – December 2019)}

The optimal values from the data collected from this period included alpha set at 0.01, beta at 0.5, 20 topics, and 200 iterations. The coherence and perplexity scores from these values of optimal hyperparameters were 0.43 and -7.87, respectively. All 20 topics were inferred from a combination of ten keywords and the weight that was assigned to each of them by the topic model (see Table \ref{tab:table3} and Appendix \ref{before}).

\begin{table}
  \caption{Major topics in the pre–pandemic period}
  \label{tab:table3}
  \resizebox{.95\columnwidth}{!}{
  \begin{tabular}{|p{2cm}|p{5cm}|p{3cm}|}
    \toprule
    \textbf{Topic Number} & \textbf{Frequent Terms} & \textbf{Inferred Topics} \\
    \midrule
    {1} & {0.138*"friends" + 0.101*"family" + 0.028*"homophobic" +
    0.017*"mom" + 0.016*"gay" + 0.013*"friend" + 0.013*"dad" + 0.012*"old" +
    0.010*"parents" + 0.009*"year"} & {Friends and Family} \\
    \hline
    {2} & {0.054*"time" + 0.033*"first" + 0.025*"sexual" + 0.020*"health" +
    0.016*"discussion" + 0.009*"sex" + 0.008*"transgender" + 0.007*"ace" +
    0.006*"asexual" + 0.006*"bought"} & {Sex education} \\
    \hline
    {3} & {0.165*"lgbt" + 0.117*"non" + 0.018*"discussion" + 0.013*"fuck" +
    0.008*"binary" + 0.007*"gay" + 0.007*"big" + 0.004*"got" + 0.004*"new" +
    0.004*"community"} & {Discussions about shared sense/notion of community
    and identity} \\
  \bottomrule
\end{tabular}
}
\end{table}

 

\subsection{Sentiment Analysis}

The initial accuracy of the RoBERTa model without finetuning was 68.5\%, with a 67.5 F1 score, 68.8 precision, and 67.0 recall. After finetuning our RoBERTa model, its accuracy was 79\%, with a precision of 79.5, recall of 79.0, and F1 score of 79.0. Comments gathered from all the subreddits both before and during the pandemic are labeled by this finetuned RoBERTa model (see Table \ref{tab:table4} and Table \ref{tab:table5}).

\begin{table}[h!]
\renewcommand{\arraystretch}{1.25} 
\centering
\small
\caption{Sentiment Composition in each subreddit in the pre–pandemic period}
\label{tab:table4}
\resizebox{\columnwidth}{!}{%
\begin{tabular}{|c|c|c|c|}
\toprule
    \textbf{Subreddit} & \textbf{\% of Positive Comments} & \textbf{\% of Negative Comments} & \textbf{\% of Neutral Comments} \\
    \midrule
    {r/lgbt} & {27.84} & {26.50} & {45.66} \\
    \hline
    {r/LGBTeens} & {21.50} & {22.90} & {55.60} \\
    \hline
    {r/ainbow} & {13.64} & {43.16} & {43.19} \\
    \hline
    {r/LGBTnews} & {9.55} & {47.39} & {43.06} \\
    \bottomrule
\end{tabular}}
\end{table}

\begin{table}[h!]
\renewcommand{\arraystretch}{1.25} 
\centering
\small
\caption{Sentiment Composition in each subreddit during the pandemic period}
\label{tab:table5}
\resizebox{\columnwidth}{!}{%
\begin{tabular}{|c|c|c|c|}
\toprule
    \textbf{Subreddit} & \textbf{\% of Positive Comments} & \textbf{\% of Negative Comments} & \textbf{\% of Neutral Comments} \\
    \midrule
    {r/lgbt} & {27.88} & {19.61} & {52.50} \\
    \hline
    {r/LGBTeens} & {22.60} & {19.61} & {57.78} \\
    \hline
    {r/LGBTQ} & {16.44} & {30.61} & {52.94} \\
    \hline
    {r/ainbow} & {16.80} & {33.50} & {49.70} \\
    \hline
    {r/LGBTnews} & {10.16} & {43.83} & {46.00} \\
    \bottomrule
\end{tabular}}
\end{table}

A shift in tone is observed from the period before the pandemic to during the pandemic. Positive comments have marginally but discernably increased, along with a marked decrease in the negative comments. Additionally, there is a noticeable increase in the percentage of neutral comments as well.

\section{Discussion}

The themes we have observed from topic modeling and patterns in sentiment analysis have been discussed in detail in this section.

\subsection{Topic Modeling}

We searched for the combination of ten keywords used to assign topic labels in the dataset, and comments with a mixture of such words were analyzed to grasp meaningful insights into what constituted within those topics. 
 
There is a stark difference noticed in the results of the topic model from the periods before the pandemic and after the pandemic. Coming out to friends and families, discussion around gendered identities and sexual orientations, drag shows and performance of gender, and sharing school experiences and relationship advice seem to be the major topics discussed in the pre–pandemic period. However, political discussions and debates on events happening worldwide increased significantly during the period of the pandemic. Members of these subReddit communities shared resources and educated other members on the history of current events and the stigma faced by LGBTQ+ communities in different parts of the world during the pandemic.

One theme that remained common across the two time periods, was the discussion of Pride Celebrations. The themes that were found in the pre–pandemic period did not die down during the pandemic. However, during the pandemic, users were found to be distressed by the implication of the legislations in their life. Therefore, these legislations and other related events took precedence and were being considered more important.

\subsubsection{Political Discussions}

Through manual evaluations of the comments, we find that during the pandemic terms like vulnerable, stress, friends, health, surgery, AIDS, and other similar words in the topic model appeared due to the difficulties faced by LGBTQ+ communities. For instance, r/LGBTnews shared a post titled, ``The Trump administration is being sued for trying to remove queer people from human rights laws." In June 2020, when the COVID–19 pandemic was well underway, the Trump administration (2017–2021) reversed the protections put in place by the previous Obama administration (2009–2017) in health care, encouraging healthcare providers to deny care to transgender people. There are also several studies that point to similar, institutionalized forms of discriminations against LGBTQ+ individuals and communities in other countries during the pandemic\cite{gil2021lesbian,wallach2020address}.
 
News headlines from around the world were shared as posts on the subreddits, particularly the subreddit, r/LGBTnews, in the course of the pandemic. By studying these, we find that the world, even while facing a global crisis, continued to discriminate against gender and sexual minorities. In fact, the discrimination and prejudice they faced was further aggravated.  One such post consisted of a headline from India, which stated that “Horrific propaganda campaign claims coronavirus is spread by talking to transgender people.” These headlines demonstrated the nature of fear, prejudice, and paranoia associated with transgender individuals and communities at the time, often through unscientific narratives suggesting that LGBTQ+ people were causing the spread of the disease. These were clearly points of serious concern and discussion for the users of the subreddits, as their presence and frequency suggests.
 
More headlines help us better understand the LDA model and why terms such as ‘Idaho’, ‘lambda’, ‘Trump’, ‘Biden’, and ‘aids’ appeared in the topic clusters. For instance, “Idaho Governor Signs The Nation’s Most Anti-Transgender Measures Into Law” is a headline from Idaho. Specific to the United States, the users had something to say about the 2021 American elections. One comment, clearly referring to the US presidential election of 2021 and conveying how many LGBTQ+ individuals thought and felt at the time, reads, “He [Joe Biden] is our only option, obviously not a great one, not even a good one, but we CANNOT afford another term of Trump”. Reddit users were, thus, spending time educating each other and making each other aware of the changes in laws and government that would affect them.
 
Further evaluations of the Reddit data revealed that users were actively discussing developments unfolding in 2020 in different parts of the world that would adversely affect their existence and that of other allies and communities like them. Singapore’s High Court upheld a law criminalizing gay sex on 30 March 2020. During the same time, the Hungarian government disallowed the legal changing of one’s gender. Both the events are intensively discussed by the users of r/LGBTnews subreddit. 
 
Another point discussed actively in these subreddits during the pandemic was representation. Not having adequate representation is a crucial point discussed actively in queer political spaces. Queer studies and activisms, as pointed out earlier, have stressed intersectional aspects of identity in recent years, and LGBTQ+ users on these subreddits believe that through representation, people will become more accepting towards them. Previous research has also emphasized the importance of representation, how issues with stereotyping and the invisibilizing of many groups within LGBTQ+ communities can be addressed with more and better representation\cite{trivette2023views,mcinroy2017perspectives}.

\subsubsection{Health}

Our analysis reveals that the prominent presence of the words “conversion” and “therapy” during the pandemic in the LDA model refers respectively to Sex Reassignment Surgery (SRS or more widely known as Gender–affirming Surgery) by transgender and non–binary individuals wishing to transition and the HIV/AIDS epidemic (1981 to Present). As mentioned above, gender–affirming surgery was one of many medical procedures that had to take the backseat in the immediate urgencies of the pandemic. However, this created more complications for transitioning individuals, leading to hormonal imbalances and increased anxiety during this period of isolation\cite{stevens2021natural}. The discussion of SRS in the subreddits during the pandemic is associated with these uncertainties and problems. Furthermore, the COVID–19 pandemic also brought back bad experiences and memories associated with the HIV epidemic, when LGBTQ+ communities faced extreme discrimination in many aspects of their life, including housing, public accommodation, and most importantly, health care, due to the stigma surrounding them.
 
The discussion around the HIV epidemic can be identified from posts titled “Isolation and HIV memories hit LGBT+ elderly hard in lockdowns”. In another post LGBTQ+ youth were seen taking advice from older generations on combating discrimination during a crisis. The reason for comparing the HIV epidemic and the COVID–19 pandemic could be attributed to the widespread and frequently fatal nature of both events, as well as the specific nature of the discrimination that LGBTQ+ communities experienced then and still face today, as recent research has also pointed out\cite{wenham2020women,logie2020we}. 
 
Along with this, conversations around hormone replacement therapy were also visible. One such comment offering advice was, “It's also okay to not be ready for HRT or surgery or even know if you want it or not when you haven't even come out yet”. Comments have also been found where users sought advice post–surgery, thus showing how community members actively helped each other out using the Reddit platform.

\subsubsection{Family, Friends, and Relationships}

Pre–pandemic results of the topic model shows a high usage of the terms rant, crushes, advice, and discussion because posts on the subreddits were marked by the Reddit users as such to categorize their posts. Through their posts, these Reddit users asked for advice as to “how do I come out?” or “idk if I am a lesbian or bisexual”. Moreover, they needed advice to determine the course of action for coming out to their family and friends. It is quite evident that these LGBTQ+ communities found love and support through their Reddit friends. This empowering connection within online communities of vulnerable groups has also been found by previous research\cite{fish2021sexual,jenzen2014make}.
 
Under the family and friends tag, one could find posts such as:
 
“I told my mom when I thought I was bi...she just said that it was a phase and she does not ever want to go to a lesbian wedding...My family is extremely homophobic...I don't know how she would react, and I just need a little advice and support. :)”.
 
The thought and introspection that goes into the difficult and even potentially dangerous process of coming out is conveyed vividly in these comments. We found that users considered coming out to be a process that demanded patience and thought and needed to be done when one was feeling safe and when their life goals would not be hindered by an open declaration of their identity. It is well-established that any public space can suddenly turn unsafe when an individual’s sexual or gender identity is questioned, and that this can lead to violence and harassment\cite{russell2021promoting,kosciw20202019}.

Intimate relationships have also emerged as a topic from the pre–pandemic period. LGBTQ+ intimate relationships suffer from numerous pressure points, but can also be the most crucial spaces for sexuality and gender to be deeply affirmed. The pressures from society to have a “normal” relationship, and guilt from disappointing parents lead to added stress in a relationship. In the case of moving out of their homes, the dependence of LGBTQ+ individuals on their partner increases manifold. They are emotionally, sexually and sometimes even financially dependent on their partner\cite{shah2015no}.
 
Schools were a place where many users described exploring their sexualities. However, the already challenging process of coming out was made much more difficult during the pandemic due to the different isolationary demands of lockdown mechanisms in several countries. Forced proximity to hostile family or community members in the absence of financial autonomy added another layer of intensification of these issues. Many LGBTQ+ individuals on the subreddits were forced to go back into the closet because leaving home during the pandemic was not feasible. What made such situations worse was that even if they could leave their homes, they found that public infrastructure, such as shelters, wherever these were available, were saturated and simply could not take more people in\cite{gil2021lesbian,konnoth2020supporting,haworth2023no}. There is more evidence of this in the data we collected from the subreddits, where the users advised individuals not accepted at home to not rely on public infrastructure and, instead, create their own network of supportive friends during the pandemic. Moreover, users advised younger people to lie to their parents until they could move out of the family home.
 
Research suggests that an unsupportive family frequently becomes an individual’s biggest source of dysphoria\cite{halliwell2018psychological,eisenberg2020family}. Our manual analysis of the results of the topic model both from during and pre–pandemic data revealed experiences of many users who had experienced adversity from their parents when they chose to dress differently and expressed that these actions were their biggest source of dysphoria.

\subsubsection{Pride Celebrations}

Pride celebrations emerged as a topic both in the pre–pandemic period and during the pandemic. During the pre–pandemic period, these celebrations helped people meet and connect with a community that was supportive and empowering. The cancellation of Pride celebrations during the pandemic was particularly concerning for members of the LGBTQ+ communities, who had reported feeling isolated and alone. Pride celebrations also act as fund-raising opportunities for legal services, health care, and other facilities in many countries. Due to the pandemic, they were either held online or were not held at all\cite{konnoth2020supporting}. Data from the subreddits revealed that this inability to conduct Pride celebrations as they usually are led to an aggravation of the economic as well as psychological and mental challenges faced by these communities. The love for Pride celebrations is also clear, as we can see from this post:
 
“I guess I just want to rant...I’m worried that there won’t be a pride parade this year...It’s one of the only places I can see some of my friends...I hope they’ll at least postpone it instead of cancelling it completely”.
 
Another post read, “LGBTQ Groups Team Up For Online 'Global Pride' Amidst Coronavirus Crisis,” thereby providing hope and determination to other users in the subreddit community to keep working towards maintaining connections and support networks during the pandemic, using whatever resources they could muster.

\section{Sentiment Analysis}

A clear pattern emerges from the results of our sentiment analysis. During a period of aggravated hostility and fear during the pandemic, the subreddits offered a safe space as indicated from the marginal but perceptible increase in positive comments. With a marked decrease in negative comments, we find that these Reddit communities are empowering spaces, because it is not that people do not express negative experiences or feelings. Indeed, negative experiences and feelings persist, and this fact aligns with what has been already documented of aggravated hostility and fears. But what is striking here is that the atmosphere of support that negative experiences and feelings brought forth, for other users offered support and positivity in response to them. In fact, it is not an exaggeration to argue that these subreddit platforms became akin to surrogate or alternative families or friends during the difficult pandemic period. Users described this phenomenon as the creation of a “little bubble of friends”, referring to the necessary creation of isolated “bubbles” to ensure that the COVID–19 contagion was contained, while also maintaining healthy social and human contact within real or virtual communities.
 
We also noticed an increase in neutral sentiments during the pandemic, which could indirectly mean an increase in negative comments. Given that a Reddit community is moderated through its own set of rules, any interaction in the form of post or comment that does not meet community guidelines is removed by the bot. RoBERTa marked comments such as [deleted], [removed by bot], and others neutral. Therefore, an increase in neutral sentiments could be due to an increase in hostile or irrelevant comments, which would suggest an increase in negative comments that target LGBTQ+ communities. Another reason could also be an obvious increase in activity as a result of restrictions on access to public spaces during the pandemic in different locations. A lot of such comments in this increased traffic cannot be classified as positive or negative, for these are not adding substantially to the content or discourse of the post and can be in the form of emoticons or one–worded replies such as ``LOL".
 
However, an increase in neutral comments is not an indication that most people had neutral sentiments regarding the LGBTQ+ topic\cite{aldinata2023sentiments} or preferred to ignore the issue entirely\cite{hoiriyah2023sentiment}. Neutral comments in themselves contained a lot of information regarding the experiences of the LGBTQ+ communities and the topics discussed. An example of a neutral comment from r/ainbow subreddit is, “Don't bring up privilege when you clearly have NO UNDERSTANDING OF WHAT IT MEANS. I have white privilege and cis privilege. Being gay doesn't erase that..."
It is evident from this example that important facts, with distinct positive or negative implications, that are being shared and discussed in these communities are marked neutral, but these demonstrate very clearly that users are interested in LGBTQ+–related events, e.g. changes in law, lived experiences during the pandemic, the process of coming out, or advice on surgery.

\section{Conclusion}

The Reddit platform gave us a means to understand the everyday experiences of LGBTQ+ users during and before the pandemic using posts and comments published by them. With huge amounts of data, it is impossible to analyze each one qualitatively. That is where topic modeling and sentiment analysis on datasets retrieved from before the pandemic and during the pandemic proved to be useful in our research. Our results show that there was a shift in both theme and affect observed from pre–pandemic to during the pandemic period. While fundamental issues that rise above social, political, and practical realities—such as abstract issues of self-definition and self-representation—continue to be discussed in these subreddits, there is a stark difference in other respects. The discussions in the pre–pandemic period largely focused on the process of coming out to friends and family members, experiences at work places, and seeking advice and support on a varying number of topics, which included exploring sexualities. However, attention during the pandemic turned to political events worldwide, as users anxiously and indignantly discussed world affairs, particularly those that actively sought to discriminate against LGBTQ+ groups and communities. The precarity of being “different” had increased during the pandemic. 

In addition to this, the results from sentiment analysis show us an increase in positive and neutral sentiments with a decrease in negative sentiment from pre–pandemic to during the pandemic period. We found that neutral comments contained a lot of information regarding the experiences of the LGBTQ+ communities and the topics discussed. An increase in neutral comments was not an indication of ignorance of LGBTQ+ issues but rather could be due to an increase in hostile comments targeting LGBTQ+ individuals or simply one–worded replies increasing traffic on the subreddits. Coupled with the results from the quantitative analysis, the manual analysis of random comments and posts done by us indicates that the Reddit communities provided a safer, more comforting alternative to the real world during the pandemic to LGBTQ+ users. Thus, despite the very real and tangible concerns associated with questions of data privacy, surveillance, and the misuse of AI on/using social media data, some social media platforms can and do provide space for community and belonging for vulnerable groups. 
 
We find that these subreddits act as a support network for members belonging to these LGBTQ+–centric subreddits as well as a resource for outsiders and allies interested in learning about such marginalized identities and lived experiences. They are found by chance or actively created by community members and users. While it is clear that these online communities cannot be a complete substitute for physical social and public spaces that offer community and support, it is clear that they rose to the challenge and offered such a platform for online LGBTQ+ individuals and communities during a desperate crisis. In so doing, these subreddits supply ways to imagine non–cis–heteronormative spaces for both people who belong to such communities as well as allies or others who are associated with them to exist together and articulate a meaningful discourse on a range of questions, concerns, and lived realities. This research provides insights into the social dynamics of LGBTQ+ communities that further help us better understand the changes in emotional health and the difficulties faced by the communities during the pandemic. Youngsters and adults alike were found seeking advice on various subjects during both our chosen time periods, along with discussing policies and other events that affected them. These insights from our research can better inform policies, support services, and crisis management strategies to become more inclusive by tailoring to LGBTQ+ individuals.

\section{Limitations and Future Work}

Since our research focuses on understanding the conversations that played out in a semi-anonymous social media platform, we were not able to examine in depth how other identities—such as race, caste, class, religion, and/or disability—modified and shaped the lived experience of LGBTQ+ users during the pandemic. We have tried our best to account for it in our manual analysis, but the lack of more specific information about the users’ backgrounds limits the scope of our insights. Furthermore, Reddit itself cannot be taken as a comprehensive or universal representation of LGBTQ+ experiences during the pandemic, and our study does not claim to offer such an account. Therefore, another extension to this project can consider diving deeper into qualitative analysis and conducting interviews with LGBTQ+ individuals in a defined regional context to understand how they identify and classify their discussions, concerns, attitudes, and experiences as well as gather insights on other, hitherto unexplored aspects of their lives in relation to the pandemic.
 
Furthermore, our transformer–based sentiment analysis model RoBERTa can be further trained on larger amounts of data to enhance it for the fine–tuning process, improve its ability to generalize and handle edge cases more accurately by understanding the context. Such training would be especially valuable in the context of data that is specific to the lived experiences of marginalized groups and communities, thereby enabling the building of an even more sophisticated model, sensitive to social and historical realities.

Finally, the scope of this research covers the data from before and during the pandemic period. Gathering data from after the pandemic will further enhance understanding of whether and, if so, how the COVID–19 pandemic had a lasting impact on the lives of LGBTQ+ people.

\bibliographystyle{ACM-Reference-Format}
\bibliography{sample-base}

\appendix

\section{Topic Modeling Results}

\subsection{Time Period I (10 March 2020 – 31 August 2021)}\label{during}

From the topic models trained on the optimal hyperparameters, we interpreted the topics based on a combination of 10 keywords and the weight assigned to each (see Table \ref{tab:table6}).

\begin{table}
  \caption{Major topics during the pandemic period}
  \label{tab:table6}
  \resizebox{.95\columnwidth}{!}{
  \begin{tabular}{|p{2cm}|p{6cm}|p{4cm}|}
    \toprule
    \textbf{Topic Number} & \textbf{Frequent Terms} & \textbf{Inferred Topics} \\
    \midrule
    {1} & {0.096*"first" + 0.029*"news" + 0.022*"community" + 0.020*"time" +
    0.017*"history" + 0.016*"child" + 0.015*"country" + 0.014*"state" +
    0.012*"becomes" + 0.012*"ever"} & {History and current news} \\
    \hline
    {2} & {0.095*"us" + 0.020*"coming" + 0.020*"homophobic" + 0.017*"white"
    + 0.016*"lgbt" + 0.013*"party" + 0.012*"home" + 0.012*"queer" +
    0.010*"story" + 0.010*"may"} & {Discussing and/or coming to terms with
    queer and other identities} \\
    \hline
    {3} & {0.123*"pride" + 0.031*"rainbow" + 0.029*"students" +
    0.029*"month" + 0.026*"flag" + 0.019*"protest" + 0.016*"march" +
    0.014*"flags" + 0.013*"lgbtq" + 0.012*"university"} & {Pride
    Celebrations} \\
    \hline
    {4} & {0.062*"gender" + 0.034*"kids" + 0.033*"trans" + 0.024*"non" +
    0.019*"binary" + 0.016*"drag" + 0.015*"birth" + 0.014*"change" +
    0.012*"parents" + 0.012*"new"} & {Understanding transgender and other
    non-binary identities} \\
    \hline
    {5} & {0.130*"transgender" + 0.047*"trans" + 0.044*"bill" + 0.036*"ban"
    + 0.028*"youth" + 0.028*"sports" + 0.027*"women" + 0.020*"athletes" +
    0.016*"house" + 0.015*"state"} & {Political discussions and debates with
    an interest in sports} \\
    \hline
    {6} & {0.239*"gay" + 0.049*"man" + 0.025*"openly" + 0.024*"comes" +
    0.020*"mayor" + 0.019*"first" + 0.018*"san" + 0.014*"boy" +
    0.014*"found" + 0.013*"latest"} & {News on the achievements of the
    LGBTQ+ communities} \\
    \hline
    {7} & {0.058*"lgbt" + 0.039*"support" + 0.027*"children" + 0.025*"uk" +
    0.023*"government" + 0.018*"sign" + 0.015*"republican" +
    0.013*"lawmaker" + 0.012*"even" + 0.012*"conservative"} & {Political
    discussions} \\
    \hline
    {8} & {0.036*"across" + 0.032*"anyone" + 0.030*"article" +
    0.028*"english" + 0.027*"seen" + 0.025*"paper" + 0.025*"morning" +
    0.025*"researching" + 0.025*"appalled" + 0.025*"stumbled"} &
    {Discussions of current news} \\
    \hline
    {9} & {0.057*"new" + 0.042*"lesbian" + 0.026*"sexual" + 0.021*"video" +
    0.019*"right" + 0.016*"face" + 0.016*"bisexual" + 0.015*"attack" +
    0.013*"launches" + 0.011*"gets"} & {Unwelcome attention or
    representation of gay people} \\
    \hline
    {10} & {0.070*"trans" + 0.034*"school" + 0.029*"year" + 0.022*"act" +
    0.019*"rowling" + 0.018*"jk" + 0.018*"transphobic" + 0.014*"stop" +
    0.013*"laws" + 0.013*"film"} & {Homophobic and transphobic popular
    culture} \\
    \hline
    {11} & {0.140*"lgbt" + 0.022*"help" + 0.022*"polish" + 0.021*"fight" +
    \textquotesingle0.019*"campaign" + 0.019*"activists" + 0.015*"back" +
    0.015*"called" + 0.014*"speech" + 0.012*"hate"} & {LGBTQ+--related
    debates and activisms across the world} \\
    \hline
    {12} & {0.108*"anti" + 0.100*"lgbtq" + 0.027*"group" + 0.022*"christian"
    + 0.022*"hate" + 0.021*"could" + 0.020*"religious" + 0.016*"violence" +
    0.015*"new" + 0.015*"groups"} & {Violence against LGBTQ+ communities} \\
    \hline
    {13} & {0.123*"people" + 0.040*"gay" + 0.039*"pandemic" + 0.036*"trans"
    + 0.034*"going" + 0.031*"around" + 0.026*"homophobes" +
    0.025*"amsterdam" + 0.025*"deadly" + 0.023*"spitting"} & {Experiences of
    LGBTQ+ communities with homophobic individuals} \\
    \hline
    {14} & {0.018*"lgbt" + 0.018*"free" + 0.015*"dies" + 0.014*"finally" +
    0.013*"tv" + 0.012*"claims" + 0.011*"center" + 0.011*"justice" +
    0.011*"election" + 0.011*"vote"} & {Political discussions and debates
    around representation} \\
    \hline
    {15} & {0.068*"police" + 0.068*"men" + 0.042*"arrested" + 0.029*"mexico"
    + 0.025*"release" + 0.024*"beach" + 0.022*"officers" + 0.022*"crowd" +
    0.021*"forces" + 0.021*"kissing"} & {Discussions around current news in
    relation to state apparatuses (police)} \\
    \hline
    {16} & {0.064*"biden" + 0.053*"amp" + 0.039*"law" + 0.033*"lgbtq" +
    0.024*"two" + 0.024*"hungary" + 0.023*"joe" + 0.022*"calls" +
    0.021*"day" + 0.018*"president"} & {LGBTQ+--related political
    discussions worldwide} \\
    \hline
    {17} & {0.072*"says" + 0.039*"trump" + 0.032*"people" + 0.022*"straight"
    + 0.018*"married" + 0.017*"couples" + 0.017*"said" + 0.016*"getting" +
    0.013*"come" + 0.012*"poland"} & {Debates around same--sex marriage} \\
    \hline
    {18} & {0.048*"black" + 0.031*"coronavirus" + 0.026*"say" +
    0.025*"virginia" + 0.024*"love" + 0.023*"lives" + 0.018*"lgbtq" +
    0.018*"new" + 0.017*"show" + 0.017*"years"} & {Stigma against the
    community during the pandemic} \\
    \hline
    {19} & {0.068*"sex" + 0.067*"court" + 0.035*"legal" + 0.031*"marriage" +
    0.030*"therapy" + 0.028*"conversion" + 0.026*"supreme" + 0.022*"lambda"
    + 0.019*"trump" + 0.017*"health"} & {Debates around same--sex marriages
    and conversion therapy during the Trump administration} \\
    \hline
    {20} & {0.108*"lgbtq" + 0.095*"rights" + 0.039*"woman" + 0.030*"covid" +
    0.029*"death" + 0.026*"rep" + 0.025*"john" + 0.025*"early" +
    0.024*"community" + 0.023*"lewis"} & {Stigma against the community
    during the pandemic} \\
  \bottomrule
\end{tabular}
}
\end{table}

\subsection{Time Period II (January 2019 – December 2019)}\label{before}

All 20 topics were inferred from a combination of ten keywords and the weight that was assigned to each of them by the topic model (see Table \ref{tab:table7}).

\begin{table}
  \caption{Major topics in the pre–pandemic period}
  \label{tab:table7}
  \resizebox{.95\columnwidth}{!}{
  \begin{tabular}{|p{2cm}|p{6cm}|p{4cm}|}
    \toprule
    \textbf{Topic Number} & \textbf{Frequent Terms} & \textbf{Inferred Topics} \\
    \midrule
    {1} & {0.138*"friends" + 0.101*"family" + 0.028*"homophobic" +
    0.017*"mom" + 0.016*"gay" + 0.013*"friend" + 0.013*"dad" + 0.012*"old" +
    0.010*"parents" + 0.009*"year"} & {Friends and Family} \\
    \hline
    {2} & {0.068*"discussion" + 0.059*"anyone" + 0.034*"sexuality" +
    0.031*"confused" + 0.030*"someone" + 0.028*"else" + 0.027*"talk" +
    0.024*"wanna" + 0.013*"questioning" + 0.011*"chat"} & {Exploring
    sexuality and seeking advice} \\
    \hline
    {3} & {0.054*"time" + 0.033*"first" + 0.025*"sexual" + 0.020*"health" +
    0.016*"discussion" + 0.009*"sex" + 0.008*"transgender" + 0.007*"ace" +
    0.006*"asexual" + 0.006*"bought"} & {Sex education} \\
    \hline
    {4} & {0.061*"guys" + 0.021*"hey" + 0.019*"oh" + 0.013*"god" +
    0.008*"lmao" + 0.008*"everything" + 0.008*"thank" + 0.007*"together" +
    0.006*"title" + 0.006*"christian"} & {Coming out to families} \\
    \hline
    {5} & {0.125*"relationships" + 0.043*"boyfriend" + 0.024*"relationship"
    + 0.021*"girlfriend" + 0.021*"trans" + 0.016*"discussion" +
    0.012*"dating" + 0.012*"bf" + 0.011*"got" + 0.009*"gf"} & {Intimate
    relationships} \\
    \hline
    {6} & {0.071*"friend" + 0.063*"best" + 0.026*"idk" + 0.024*"well" +
    0.021*"went" + 0.013*"outed" + 0.013*"happened" + 0.011*"maybe" +
    0.007*"sent" + 0.007*"accidentally"} & {Revealing identity to friends
    with/without consent} \\
    \hline
    {7} & {0.053*"video" + 0.034*"question" + 0.028*"bisexual" +
    0.021*"discussion" + 0.011*"song" + 0.008*"monday" + 0.007*"gay" +
    0.006*"real" + 0.006*"great" + 0.005*"music"} & {Cultural
    representations of sexuality and gender identity} \\
    \hline
    {8} & {0.152*"picture" + 0.029*"made" + 0.014*"gay" + 0.013*"thought" +
    0.010*"new" + 0.010*"found" + 0.009*"meme" + 0.009*"post" + 0.009*"day"
    + 0.009*"saw"} & {Sharing relatable memes and jokes} \\
    \hline
    {9} & {0.173*"crushes" + 0.074*"crush" + 0.025*"straight" + 0.023*"girl"
    + 0.023*"guy" + 0.013*"friend" + 0.012*"boy" + 0.012*"gay" +
    0.011*"cute" + 0.009*"date"} & {Discussions about crushes} \\
    \hline
    {10} & {0.131*"rant" + 0.050*"discussion" + 0.042*"gay" + 0.032*"people"
    + 0.023*"feel" + 0.015*"really" + 0.012*"want" + 0.011*"bi" +
    0.010*"hate" + 0.009*"straight"} & {Exploring sexuality} \\
    \hline
    {11} & {0.041*"please" + 0.032*"discussion" + 0.022*"name" +
    0.021*"help" + 0.017*"hi" + 0.013*"gsa" + 0.009*"hello" + 0.008*"tips" +
    0.008*"figure" + 0.007*"hair"} & {Seeking general advice} \\
    \hline
    {12} & {0.077*"love" + 0.022*"shit" + 0.018*"lonely" + 0.017*"sad" +
    0.017*"school" + 0.013*"high" + 0.009*"fucking" + 0.008*"experience" +
    0.007*"holy" + 0.006*"discussion} & {School experience and expressing
    feelings} \\
    \hline
    {13} & {0.040*"gender" + 0.028*"discussion" + 0.012*"work" +
    0.010*"identity" + 0.008*"questions" + 0.007*"good" + 0.007*"moment" +
    0.006*"watching" + 0.005*"church" + 0.004*"crisis"} & {Seeking advice on
    sexuality and gender identity in public contexts} \\
    \hline
    {14} & {0.018*"may" + 0.017*"queer" + 0.009*"stupid" + 0.007*"thanks" +
    0.006*"trouble" + 0.005*"perfect" + 0.005*"nsfw" + 0.005*"discussion" +
    0.004*"area" + 0.004*"online"} & {Sharing relatable posts} \\
    \hline
    {15} & {0.083*"pride" + 0.032*"happy" + 0.030*"picture" + 0.022*"month"
    + 0.016*"discussion" + 0.016*"flag" + 0.009*"got" + 0.009*"flags" +
    0.009*"go" + 0.008*"first"} & {Pride Celebrations} \\
    \hline
    {16} & {0.165*"lgbt" + 0.117*"non" + 0.018*"discussion" + 0.013*"fuck" +
    0.008*"binary" + 0.007*"gay" + 0.007*"big" + 0.004*"got" + 0.004*"new" +
    0.004*"community"} & {Discussions about shared sense/notion of community
    and identity} \\
    \hline
    {17} & {0.021*"gay" + 0.019*"discussion" + 0.018*"find" +
    0.013*"closeted" + 0.011*"teen" + 0.009*"college" + 0.009*"article" +
    0.009*"small" + 0.007*"genders" + 0.007*"rights"} & {Coming out process
    during teen years} \\
    \hline
    {18} & {0.109*"need" + 0.108*"help" + 0.088*"advice" +
    0.038*"discussion" + 0.010*"discord" + 0.008*"bit" + 0.007*"needed" +
    0.006*"situation" + 0.006*"vent" + 0.005*"pls"} & {Seeking advice} \\
    \hline
    {19} & {0.026*"discussion" + 0.019*"see" + 0.013*"last" + 0.013*"night"
    + 0.008*"problem" + 0.008*"amazing" + 0.007*"school" + 0.006*"closet" +
    0.006*"oof" + 0.006*"social"} & {Discussions around school and the
    process of coming out} \\
    \hline
    {20} & {0.190*"coming" + 0.050*"came" + 0.048*"come" + 0.026*"bi" +
    0.024*"parents" + 0.013*"friend" + 0.012*"discussion" + 0.011*"story" +
    0.010*"mom" + 0.009*"wish"} & {Coming out to friends and families} \\
  \bottomrule
\end{tabular}
}
\end{table}

\end{document}